\documentclass[10pt,letterpaper]{article}
\usepackage{opex3}
\usepackage[title,titletoc,toc]{appendix}
\usepackage{graphics}
\usepackage{cite}
\usepackage{float}
    \begin{document}
\title{Combined action of the bound-electron nonlinearity and the tunnel-ionization current
in low-order harmonic generation in noble gases}
    \author{Usman Sapaev, Anton Husakou and Joachim Herrmann $^{*}$}
    \address{Max-Born-Institute for Nonlinear optics and Fast Spectroscopy,
    Max-Born-Str. 2a, Berlin D-12489, Germany}
    \email{$^*$jherrman@mbi-berlin.de}
    {We study numerically low-order harmonic generation
    in noble gases pumped by intense femtosecond laser pulses in the tunneling ionization regime.
    We analyze the influence of the phase-mismatching on this
    process,
    caused by the generated plasma, and study in dependence
    on the pump intensity the origin of harmonic generation
    arising either from the bound-electron nonlinearity or
    the tunnel-ionization current. It is shown
    that in argon the optimum pump intensity of about 100 TW/cm$^2$ leads
    to the maximum efficiency, where the main contribution to low-order harmonics originates
    from the bound-electron third and fifth order
    susceptibilities, while for intensities higher than 300 TW/cm$^2$ the
    tunnel-ionization current plays the dominant role.
    Besides, we predict that VUV pulses at 133 nm can be generated
    with relatively high efficiency of about $1.5\times10^{-3}$ by 400 nm pump pulses.}

    \section{Introduction}
    Low-order harmonic generation (LOHG) in gases pumped by ultrashort near-IR laser
    pulses is an important technique to generate ultraviolet (UV) and vacuum
    ultraviolet (VUV) femtosecond pulses for a wide variety of applications,
    in particular, for time-resolved spectroscopy of many molecules, clusters or biological
    specimens and for material characterization \cite{Hertel-2006}. The use of noble gases as nonlinear medium
    instead of solid-state crystals is a preferable way to avoid strong dispersion,
    bandwidth limitations, low damage thresholds and strong absorption below 200 nm.
    In particular, by using different gases UV and VUV pulses with a duration down to 11 fs
    have been generated by third \cite{Backus-1996,Trushin-2007}
    and fifth harmonic \cite{Kosma-2008} conversion.
       Similar to other nonlinear frequency conversion processes, the efficiency of LOHG in gases
    is usually relatively low in practice. This is mainly
    caused by two factors: first, low conversion results from relatively
    small values of the third (and higher) order susceptibilities compared to crystalline
    media. A second problem is the realization of phase matching, which can be partially solved
    for various frequency transformation processes,  e.g., by using the anomalous
    dispersion of hollow-core fibers \cite{Durfee97,tzankov07,babushkin08},
    modulated hollow-core waveguides \cite{Paul}, a modulated third order nonlinearity by ultrasound
   \cite{sapaev2012} or by noncollinear four-wave mixing \cite{Ghotbi-2010,Ghotbi-2013}.

       In the intensity range below the ionization threshold the efficiency of
       frequency transformation increases with increasing pump intensity.
       However, as soon as the intensity rises above the ionization threshold,
       different additional processes play a role leading to a more complex dynamics.
       On the one hand the effective third-order nonlinearity decreases,
       since $\chi^{+(3)}$ of an
    ionized gas is lower than $\chi^{(3)}$ of a corresponding neutral gas
    \cite{Bree2011}. On the other hand harmonics of the fundamental
    frequency emerge due to the ionization of
the atoms and the interaction of the freed electrons with the
intense pump field. The majority of studies of harmonic generation
  have been performed for relatively high orders
  of harmonics much in excess of the ionization
  potential which is well described by the three-step
  process of ionization, acceleration in the continuum
   and recombination with the parent ion
   (see e.g. \cite{Krausz}). Much less
 studied is an additional, physically different,
          mechanism of optical harmonic generation. In the regime of tunneling
          ionization the density of ionized electrons shows extremely
          fast,
          nearly stepwise increases at every half-cycle of the laser field.
          This stepwise modulation of the tunnel ionization current induces
          optical harmonic generation \cite{Brunel-1990}, which arises in the
          first stage of ionization and not in the final recombination stage in
          the three-step model. As shown in \cite{Burnett-1995}, the emission
           of the lowest harmonics up to about 9 are accounted for with the tunnel
           ionization current while higher orders are attributed with the
           recombination process. Further theoretical studies of this process has been
           published in \cite{Vanin-1993,Serebryannikov-2009}.
           Note that the generation
of THz pulses by two-color femtosecond pulses is also intrinsically
connected to the optically induced step-wise increase of the plasma
density due to tunneling ionization
\cite{Kim_2007,Babushkin2010,Babushkin_2011}.

           To date only few direct experimental observations of harmonic generation or
           frequency mixing due to the modulation of the tunnel ionization current has
           been reported \cite{Siders-2001,Verhoef-2010}. On the other hand the generation
           of third harmonics with efficiencies up to the range of $10^{-3}$ in a noble
           gas with pump intensities significantly larger than necessary for ionization has
           been reported in \cite{Backus-1996,Siders-1996}.
           Detections of this type of nonlinear response in the regime with intensites above
           the ionization threshold requires better understanding of the complex
           dynamics and insight into the
           competition of harmonic generation originating from atomic or ionic
           susceptibilities of bound states and the tunneling ionization current. The present paper is devoted to a theoretical study of this issue.
           Since here only low orders up to 7 are considered we neglect the recombination process
           and consider only the first stage of ionization.
           At least, up to our knowledge the combined occurrence of the two mechanisms
           by the bound-electron third (and higher) order nonlinearity and the tunnel-ionization current were not
           studied before.
    \section{Fundamentals}
    Third harmonic generation (THG) in gases by
    focused beams for pump intensities below
    the ionization threshold has been studied
    theoretically already four decades ago
    \cite{Ward-1969,Bjorklund-1975}.
    In the regime of tunneling ionization besides the nonlinearity due to
    bound electron states additional processes come into play which has to be
    accounted for. In particular, the sub-cycle temporal dynamic of the laser field plays
    an essential role in the ionization process. Therefore, in the theoretical description
    the slowly-varying envelope approximation requires the solution of a complicated,
    strongly coupled system of partial differential equations
    and would result in
    increased numerical errors due to relatively short (down to $8$ fs) durations of harmonic pulses.
    Since backward propagating field components are small we can use
    the unidirectional pulse propagation equation for the description of pulse
    propagation \cite{Husakou-2001}. As will be seen later the effective propagation length
    is much smaller than the Rayleigh length,
    therefore we can neglect the diffraction term. Correspondingly
    the following basic equation for the electric field of linear polarized pulses will be used:
    \begin{eqnarray}
    \partial_{z}\hat{E}(\omega) & = & ik(\omega)\hat{E}(\omega)+i\frac{\mu_{o}\omega^{2}}{2k(\omega)}\hat{P}_{NL}(\omega)
    \end{eqnarray}
    Here $\hat{E}(\omega)$ is the Fourier transform of the electric field $E(t)$;
    $k(\omega)=cn(\omega)/\omega$ is the frequency-dependent wavenumber,
    $\omega$ is the angular frequency, $c$ is the speed of light and
    $n(\omega)$ is the frequency-dependent refractive index of the chosen
    gas;
 $\mu_{o}$ is the vacuum
    permeability. The first term on the right-hand side of Eq. (1)
    describes linear dispersion of the gas. The 
    nonlinear polarization is
    $\hat{P}_{NL}(\omega)=\hat{P}_{Bound}(\omega)+i\hat{J}_{e}(\omega)/\omega+i\hat{J}_{Loss}(\omega)/\omega$ with
    $\hat{P}_{Bound}(\omega)$ being the nonlinear polarization caused by the bound electron states, $\hat{J}_{e}(\omega)$
being
    the electron current and $\hat{J}_{Loss}(\omega)$ being the loss term due to
    photon absorption during ionization.
    The plasma dynamics is described by the free electron density
    $\rho(t)$, which can be calculated by:
    \begin{eqnarray}
    \partial_{t}\rho(t)=W_{ST}(t)(\rho_{at}-\rho(t))
    \end{eqnarray}
    where $\rho_{at}$ is the neutral atomic density; $W_{ST}(t)$ is the
    quasistatic tunneling ionization rate for hydrogenlike atoms
    \cite{Ammonov-1986}
    $W_{ST}(t)=4\omega_{a}(r_{h})^{2.5}(\left|E_a\right|/E(t))$exp$(-2r^{1.5}|E_a|/3E(t))$,
    where $E_{a}=m_{e}^{2}q^{5}/(4\pi\epsilon_{o})^{3}\hbar^{4}$,
    $\omega_{a}=m_{e}q^{4}/(4\pi\epsilon_{o})^{2}\hbar^{3}$ and $r_{h}=
    U_{Ar}/U_{h}$, $U_{h}$ and $U_{Ar}$ are the ionization potentials of
    hydrogen and argon, correspondingly;
    $P_{Bound}(t)=\epsilon_{o}\chi^{(3)}(1-\rho(t)/\rho_{at})E(t)^{3}
    +\chi^{+(3)}(\rho(t)/\rho_{at})E(t)^{3}+\chi^{(5)}(1-\rho(t)/\rho_{at})E(t)^{5}$,
    $\epsilon_{o}$ is the vacuum permittivity; $m_{e}$ and $q$ being the
    electron mass and charge, respectively; $\chi^{(3)}$ and $\chi^{(5)}$ are
    third and fifth order susceptibilities of neutral gas, correspondingly,
    while $\chi^{+(3)}$ is that of ionized
    gas. In the following we consider nearly collimated beams with diameter
    corresponding to a Rayleigh length larger than the propagation lengths. In
    addition, for these parameters the pump power is below the self-focusing power.
    Therefore, we can neglect diffraction in the numerical model.
    The transverse macroscopic plasma current $J_{e}(t)$ is determined
    by \cite{Babushkin2010}:
    \begin{eqnarray}
    \partial_{t}J_{e}(t)+\nu_{e}J_{e}(t)=\frac{q^{2}}{m_{e}}E(t)\rho(t)
    \end{eqnarray}
    where $\nu_{e}$ is the electron collision rate (for argon
    $\nu_{e}\approx5.7$ ps$^{-1}$). Finally, the ionization energy loss
    is determined by
    $J_{Loss}(t)=W_{ST}(t)(\rho_{at}-\rho(t))U_{Ar}/E(t)$.
    A critical condition for an efficient frequency transfer to harmonics is the realization of phasematching
    which for intensities larger than the ionization threshold is sensitively influenced by the plasma
    contribution to the refraction index. The change of the linear refractive index of
    argon at the maximum of the pulse intensity $I^{\prime}$ (assuming Gaussian pulse
    shape),
    owing to the formation of laser plasma with a free electron density $\rho^{\prime}$ and the
    Kerr nonlinearity, is given by (see e.g.,  \cite{BookZ,zh1}):
    \begin{eqnarray}
    n(\omega,I^{\prime},\rho^{\prime})=n_{e}(\omega,\rho^{\prime})+\Delta
    n_{Kerr}(I^{\prime},\rho^{\prime})+\Delta n_{Plasma}(\omega,\rho^{\prime})
    \end{eqnarray}
    where
    $n_{e}(\omega,\rho^{\prime})=(n^{o}(\omega)-1)(1-\rho^{\prime}/\rho_{at})+1$,
$\Delta
n_{Kerr}(I^{\prime},\rho^{\prime})=I^{\prime}[n_{2}(1-\rho^{\prime}/\rho_{at})+n_{2}^{+}\rho^{\prime}
+ I^{\prime}n_{4}(1-\rho^{\prime}/\rho_{at})]$
    and $\Delta
    n_{Plasma}(\omega,\rho^{\prime})=-q^{2}\rho^{\prime}/(2\epsilon_{o}m_{e}\omega^{2})$.

The nonlinear susceptibility $\chi ^{(3)}$ for argon is well known
from many independent measurements, while only few experimental
results exist for the higher-order susceptibilities. In
\cite{Loriot2009,Ni-2011} coincident experimental date on $\chi
^{(5)}$ for argon which also agrees (up to sign) with a theoretical
estimation \cite{BreeDis} can be found. On the other hand reported
data for $\chi ^{(7)}$ differ by orders of magnitudes.
Correspondingly, neglecting the weak frequency dependence we assume
in the following parameters $\chi ^{(3)}=3.8\times10^{-26} $
m$^2$/V$^2$ and $\chi ^{(5)}=-2.02\times10^{-47} $ m$^4$/V$^4$.

    \begin{figure}[htbp]
    \centerline{\includegraphics[width=11cm]{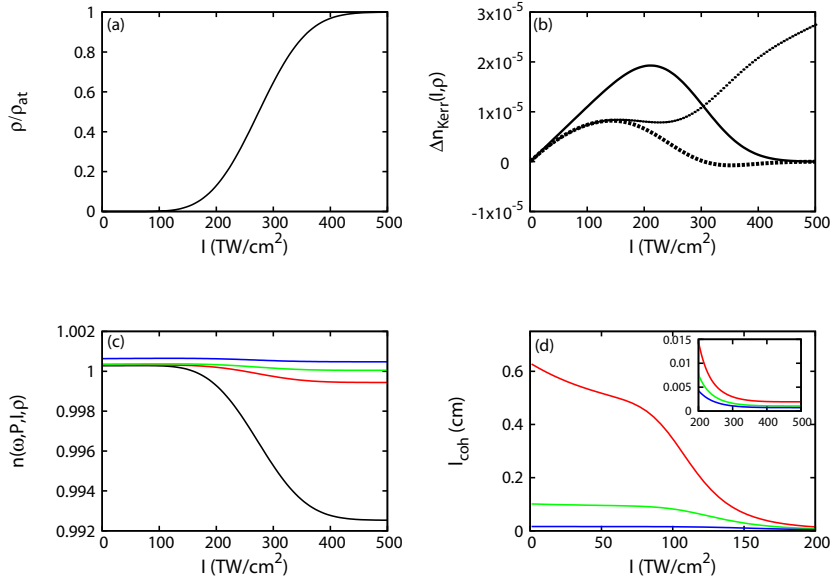}} \caption{Linear
    and nonlinear optical parameters of argon in the high-intensity regime
    (normal pressure P $=1$ atm) for a $20$ fs pulse at $800$ nm: (a)
    free electrons density at the trailing edge of the pulse normalized by the total number of
    atoms ($2.7\times10^{25}$m$^{-3}$); (b) contributions of the nonlinear
    refractive index, caused by Kerr nonlinearity, when only $n_{2}=
    1.1\times10^{-23}$m$^{2}/$W (solid), $n_{2}$ and $n_{4}=-0.36\times10^{-41}$m$^{4}/$W$^{2}$
    \cite{Loriot2009} (dashed),
     $n_{2}$, $n_{4}$ and ionized argon
    $n_{2}^{+}=0.6\times10^{-23}$m$^{2}/$W \cite{Bree2011} (dotted) are taken into account;
    (c) change of the total refractive index of pump (black), third
    (red), fifth (green) and seventh (blue), harmonics; (d) change
    of coherent length of third (red), fifth (green) and seventh
    (blue) harmonics.}
    \end{figure}

    Figure 1 shows some linear and nonlinear optical
    parameters of argon, calculated by using Eq. (4) for a $20$ fs (FWHM) pump
    pulse at $800$ nm in dependence on the pump intensity. Figure 1(a)
    shows the normalized plasma density after the pulse as a function of the pump
    intensity at the peak of the pulse. As can be seen full ionization at the trailing edge of the pulse occurs at around
    $450$ TW/cm$^2$.
    Figure 1(b) shows changes of the
    Kerr-type nonlinear refractive index contribution
    $\Delta n_{Kerr}(I,\rho)$ taking into account (i) only n$_2$ of
    neutral argon (solid curve), (ii) n$_2$ and n$_4$ of neutral
    argon (dashed line) and (iii) n$_2$, n$_4$ and n$_2^+$ of neutral
    and ionized argon (dotted line).
    Figure 1(c) shows the change of the refractive indexes of the fundamental
    frequency (black), third (red), fifth (green) and seventh (blue) harmonics. As can be seen the
    refractive indexes of the fundamental
    frequency and third harmonic are decreased down to a value smaller than unity.
    For higher intensities, the
    difference between the refractive
    indexes of the fundamental and the harmonics
    becomes larger, i.e., the generation of plasma
    electrons decreases the phase-matching length.
    This can be seen in Fig. 1(d) demonstrating the change of
    the coherent lengths $(l_{coh}=\pi/|\Delta k_{2r+1}|)$ for different
    harmonics in dependence on the pump intensity. Here the coherent
    length strongly decreases above $100$ TW/cm$^{2}$ for all
    harmonics. Below we show that this behavior appears also in our full
    numerical calculations of Eqs. (1)-(3).

Neglecting bound electron contributions and the dependence of the
pump intensity and plasma density on the propagation coordinate an
analytical solution for the electric field of the harmonic has been
derived in \cite{Brunel-1990}. If we include the bound-electron
contributions, the electric field  for the harmonics with order of
$2r+1$ can be expressed as:
\begin{eqnarray}
E_{2r+1}(z)=\sqrt{A_{P}^{2}+\delta_{1r}(A_{3}+A_{51})^{2}+\delta_{2r}A_{52}^{2}}\sin(\Delta
k_{2r+1}z/2)/(\Delta k_{2r+1}/2)\end{eqnarray}
 here $A_{p}=-\Phi k_{o}/(8\pi r(2r+1))(\omega_{pg}/\omega_{o})^{2}\left[{\rm
 {exp}(-3r^{2}/\xi)+r/(r+1){\rm
 {exp}(-3(r+1)^{2}/\xi)}}\right]E_{o}$,
 $A_{3}=3\mu_{o}c\epsilon_{o}\chi^{(3)}E_{o}^{3}\omega_{o}/8$,
 $A_{51}=15\mu_{o}c\epsilon_{o}\chi^{(5)}E_{o}^{5}\omega_{o}/2$,
 $A_{52}=5\mu_{o}c\epsilon_{o}\chi^{(5)}E_{o}^{5}\omega_{o}/32$;
 $\delta_{ij}$ is Kronecker's symbol;
    $\omega_{pg}=4\pi\rho_{at} q^{2}/m_{e}^{2}$ is the plasma frequency
    associated with the initial gas density;
    $\Phi=8\sqrt{3\pi}(\omega_{a}/\omega_{o})\xi^{1/2}\rm{exp}(-2\xi/3)$,
    $\xi=E_{a}/E_{o}$, $E_{o}$, $\omega_{o}$ and $k_{o}$ are peak
    electric field, central angular frequency and wavenumber of pump,
    respectively; $\triangle k_{2r+1}$ is wave mismatch for
    $(2r+1)^{th}$ harmonic. The term $A_{51}$ describes the contribution of
    fifth-order susceptibility $\chi^{(5)}$  to the generation of third harmonic.
     We note that the relative phase of the fields generated
    by bound-electrons and the plasma current is $\pi/2$. In the following we compare this solution with
    the numerical solutions of the full model as presented in Eqs. (1)-(3).
    \begin{figure}[htbp]
    \centerline{\includegraphics[width=8.5cm]{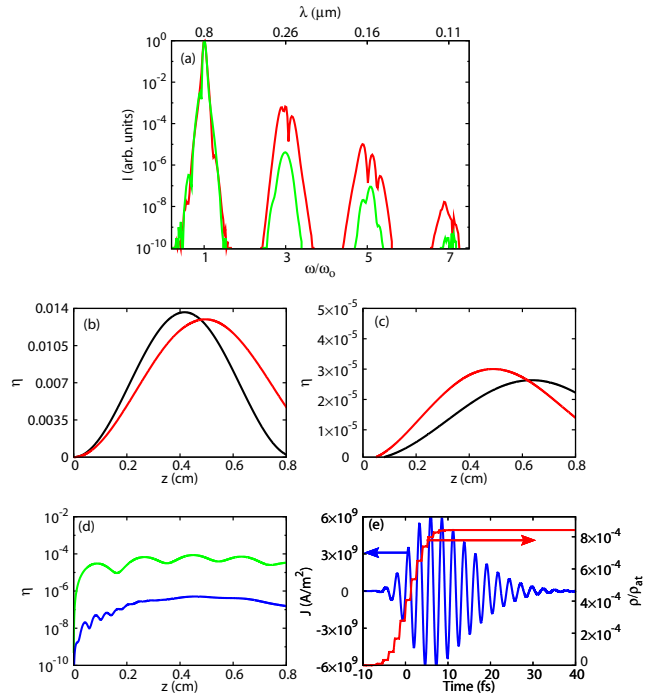}} \caption{
   Numerical
    and analytical calculations for a 20-fs transform
    limited pump pulse with a 100 TW/cm$^{2}$ peak intensity at 800
    nm:
    (a) spectrum of the output pulse, calculated numerically with
    (red) and without (green) taking $\chi^{(3)}$ and $\chi^{(5)}$ into account; (b)
    efficiency conversion of third harmonic, calculated numerically
    (red) and analytically (black) for $\chi^{(3)}\neq0$, $\chi^{(5)}\neq0$,  and
    $\chi^{(3)}=\chi^{(5)}=0$ (c); (d) efficiency conversion of fifth (green)
    and seventh (blue) harmonics; (e) normalized free electron density
    in time domain (red) and plasma current (blue).}
    \end{figure}

\section{Numerical results for 800-nm pump }

    In this chapter we present numerical solutions of Eqs. (1) to (3) using
    the split-step method with fast Fourier transformation and the fifth-order
    Runge-Kutta method for 800-nm pump pulses with a 100
    TW/cm$^{2}$ peak intensity and 20-fs (FWHM) duration.

    The spectra in Fig. 2(a) calculated with (red) and
    without (green) contribution of $\chi^{(3)}$ and $\chi^{(5)}$ predict that LOHG is
    dominated by the bound electron contributions with the third and fifth order susceptibilities,
   since with
    $\chi^{(3)}=0$ and $\chi^{(5)}=0$ two order of magnitude lower efficiencies are
    predicted. In Fig. 2(b) and 2(c) analytical (black) and
    numerical (red) results for the efficiency of third harmonic
    conversion are compared for cases, when $\chi^{(3)}\neq0$ and $\chi^{(5)}\neq0$ (b) is included
    and for $\chi^{(3)}=\chi^{(5)}=0$
    (c). Note that we calculated the efficiency  by integration of the harmonic
    spectra.

     The analytical results are calculated by Eq. (5), using
     the wave vector mismatch with a constant pump intensity and a
    plasma density taken from the input parameters. These results confirm the conclusions drawn from
    Fig. 2(a) that at the optimum intensity of about 100 W/cm$^2$ the bound-electron contribution is much larger than that
    of the tunnel ionization current. It should be noted that the maximum efficiency
    of the third harmonic of about 1.4 $\%$ appears at $0.4$ cm, which is
    approximately equal to the coherent length, as seen from Fig. 1(d). Figure 2(d) shows the efficiency of conversions to the  fifth (green) and
    seventh (blue) harmonics with maximum values of about $10^{-4}$ and $10^{-7}$.
    In Fig. 2(e) the normalized plasma current and the plasma density are presented. Note the
    steplike nature of the density profile of free electrons (red curve), which explains
    the source of the harmonic generation due to the tunnel ionization current.

       \begin{figure}[htb]
    \centerline{\includegraphics[width=11cm]{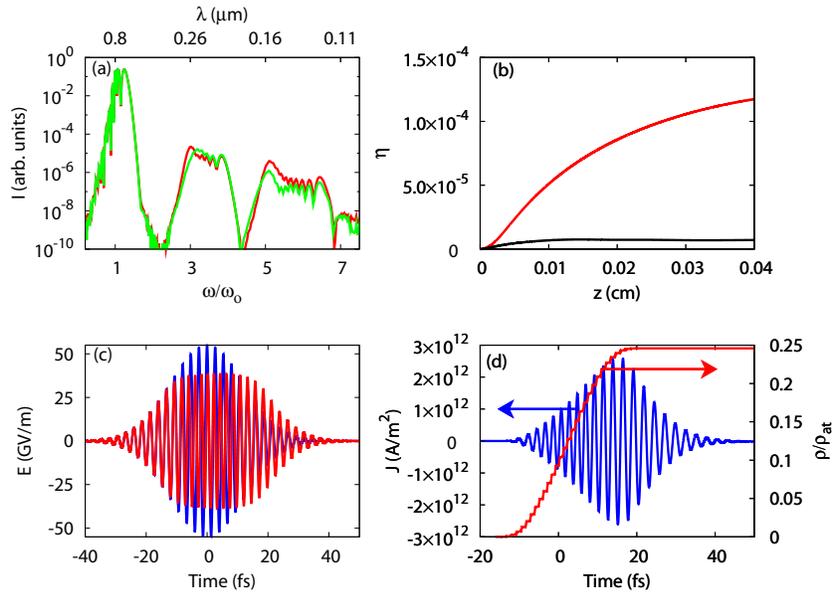}} \caption{Numerical
    calculations for a 20-fs transform limited pump pulse
    with  400 TW/cm$^{2}$ peak intensity at 800 nm: (a) spectrum of the
    output pulse, calculated with (red) and without (green) taking
    $\chi^{(3)}$ and $\chi^{(5)}$ into account; (b) conversion efficiencies of
    third
    (red) and fifth harmonics (black); (c) time profile of the pump
    at the input (blue) and output (red); (d) normalized density of free
    electron distribution (red) and plasma current (blue).}
    \end{figure}

    To study the regime where the tunnel-ionization
    current is dominant, in Fig. 3 results for LOHG are presented for a 20 fs
     pulse at 800 nm with an peak intensity of 400 TW/cm$^{2}$.
    Due to the reduced coherence length the maximum conversion
    efficiencies are smaller than for the case of lower pump intensity in Fig. 2.
    Since the efficiencies are roughly the same independent on the
    inclusion of $\chi^{(3)}$ and $\chi^{(5)}$ in the model,
    we can conclude that the tunnel-ionization current
    is the main LOHG mechanism in this case.

   Due to the high intensity significant spectral
    broadening caused by self-phase modulation can be seen. The dependence of the
    efficiencies on the propagation distance indicates pump depletion rather
    than loss of coherence, since it exhibits no maximum. Here
    pump depletion, owing
    to ionization loss, appears mainly in the pulse center, as shown in Fig. 3(c).

    In order to analyze the roles of the $\chi^{(3)}$, $\chi^{(5)}$ and  $\chi^{+(3)}$ nonlinearity
    and the tunnel-ionization current in dependence on the
    applied intensity range, we calculated the contribution to LOHG efficiencies of the two
considered
    nonlinear optical processes in a large range of pump intensities.
    Figure 4 shows the conversion efficiencies in dependence on the pump intensity
    up to the 7$^{th}$ harmonic
    from a 800-nm pump with a 20-fs duration, calculated at the coherent length of third harmonic.
    The contribution of
    $\chi^{(3)}$ and $\chi^{(5)}$ dominates up to 300 TW/cm$^{2}$, while after
    approximately 300 TW/cm$^{2}$ the plasma current (red curves)
    becomes the main source for LOHG. The green curve in Fig. 4(a) shows
    results, which includes $\chi^{+(3)}$ of the ionized gas. For high intensities up to 500 TW/cm$^2$ the efficiency of
    THG decreases down to the range of $10^{-5}$.

    \begin{figure}[htb]
    \centerline{\includegraphics[width=11cm]{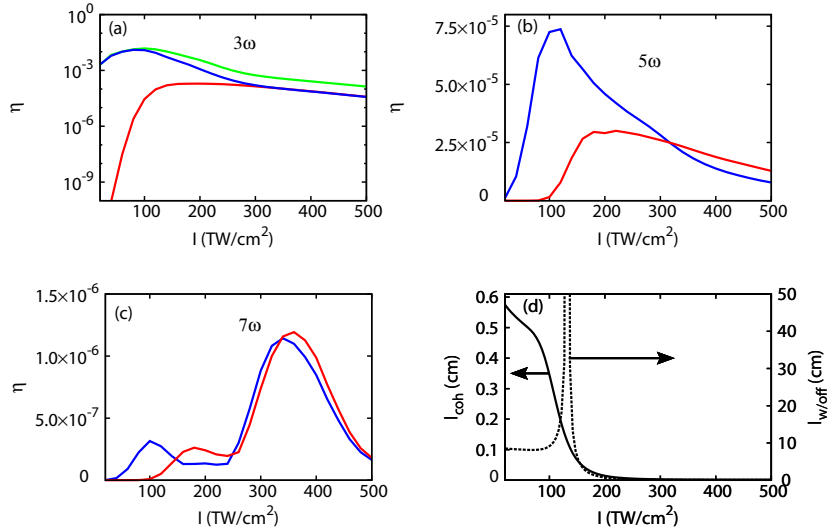}}
    \caption{Efficiency of LOHG in dependence on the pump intensity
    calculated for (a) third, (b) fifth and (c) seventh
    harmonic. Efficiencies were calculated with (blue) and without (red)
    taking $\chi^{(3)}$ and $\chi^{(5)}$ into account. Green curve in (a) shows results
    when $\chi^{(3)}$, $\chi^{(5)}$ and $\chi^{+(3)}$ were taken into account. In (d)
    the change of the coherent length of third harmonic and the length of
    its temporal walk-off from the fundamental frequency are shown.}
    \end{figure}
It seems to be surprising that even for relatively high intensities
from $150$  to $300$ TW/cm$^{2}$ the contribution of the
$\chi^{(3)}$ and $\chi^{(5)}$ process
    remains in the same order as that of the tunnel-ionization current.
    This can be explained by the fact, that full ionization only occurs at the trailing edge
    of the
    pulses, while at the leading edge the atoms are not ionized and bound-electron contributions still play a significant role.
    The length of
    temporal walk-off between the fundamental and the third harmonic is shown
    in Fig. 4(d). It is much larger than the coherent lengths,
    and therefore does not play a significant role during propagation.

    A general observation arising from the results presented above is
    that the nonlinear susceptibilities $\chi^{(3)}$ and $\chi^{(5)}$ plays an
    important role in the formation of LOHG, especially for the
    third
    harmonic. Its contribution is dominant up to a pump intensity of 300 TW/cm$^{2}$
    for argon at normal pressure. As similar behavior can be expected
    for other noble gases, although the corresponding intensities will vary dependent on the
    properties of the noble gas. The tunnel-ionization current is a main
    source for LOHG for intensities larger than approximately 300 TW/cm$^{2}$, especially for
    the fifth and seventh harmonics. As noted above, the high-intensity
    regime of pump can not support highly efficient LOHG because of the
the contribution of the ionized electrons to the refraction index
and the associated increased phase mismatch. Below 100 TW/cm$^{2}$,
the coherent length is roughly constant, but for larger intensities
it shows a sharp decrease as visible in Fig. 1(d) and
    Fig. 4(d). This establishes a range around 100 TW/cm$^{2}$ as
    optimum pump intensity for argon for the generation of the third and fifth harmonic.
        \begin{figure}[htb]
    \centerline{\includegraphics[width=11cm]{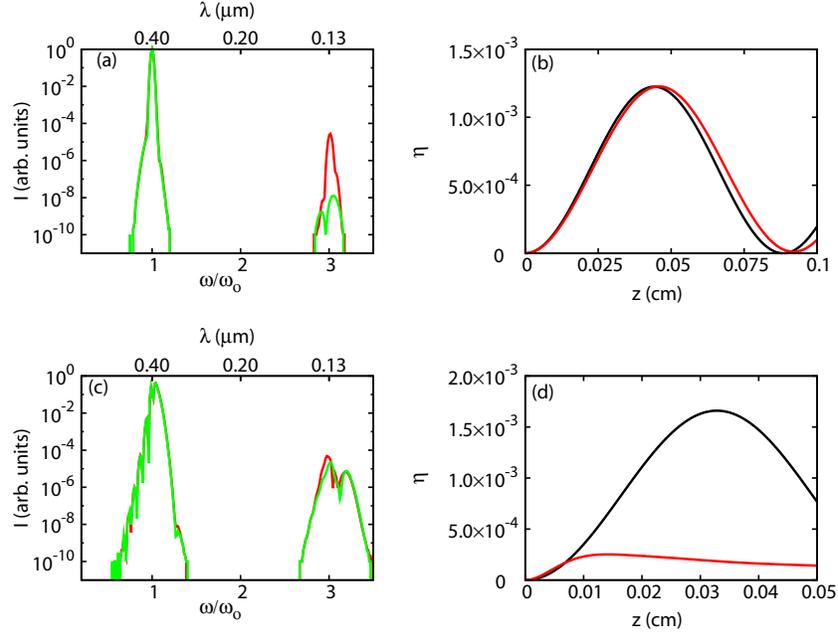}} \caption{Results
    of numerical and analytical calculations for 400 nm pump pulses with 100 TW/cm$^{2}$
    [(a),(b)] and 300 TW/cm$^{2}$ [(c), (d)]. In (a), (c)
    the spectrum of the output pulses, calculated with (red) and without
    (green) taking $\chi^{(3)}$ and $\chi^{(5)}$ are presented. In (b), (d)
    the efficiency of third harmonic, calculated numerically
    (red) and analytically (black) are shown.}
    \end{figure}
    \section{Numerical results for 400-nm pump}

    Nowadays, the generation of pump pulses at 400 nm with high energy by
    second harmonic generation in nonlinear crystals from near-IR ones is
    a standard method. Using THG with these pump pulses allows frequency conversion with
    relative high efficiency into the VUV spectral range at 133 nm.
    Figure 5 shows the results for such pump pulses with two
    different peak intensities for 100 TW/cm$^{2}$ (a, b) and for 300
    TW/cm$^{2}$ (c, d)
    and the same pulse duration of 20 fs. The coherent length for the
    third harmonic is around a $0.05$ cm
    $(0.024)$ cm for 100 (300) TW/cm$^{2}$, calculated by Eq. (4) and
    visible in Fig. 5(b) and 5(d). The tendency
    visible from Figs. 2 and 3 that for a lower intensity THG is
caused by the third and fifth order susceptibilities while for
higher intensities the tunnel ionization
    current plays the dominant role, is also observed for 400 nm
    as can be seen by comparison of Fig. 5(a) and 5(c). The maximum THG efficiency of
    about $1.5\times10^{-3}$ for a pump intensity of 100 TW/cm$^{(2)}$ [Fig.
    4(b)]
    is in the same range as in the case of a 800 nm pump pulses compare [Fig. 2(b)],
    but now a spectral transformation to the VUV range at 133 nm is realized.
    Higher harmonic orders
    above third, experience high linear loss of in the
    vacuum ultraviolet region for argon \cite{Absob} due to the strong absorption
    band below 106 nm.
    \section{Conclusions}
    In conclusion, we numerically studied the generation of low-order harmonics in
    argon in the high-intensity regime, when tunneling ionization
    takes place. The used numerical method is based on the
    unidirectional pulse propagation equation combined with the
    nonlinear response by the bound-electrons and a model
    for tunneling ionization and the associated plasma current.
    We analyzed LOHG in the regime of pump intensities up to 500 TW/cm$^{2}$
    arising either from the third- and fifth-order bound-electron
    nonlinearity or from the tunnel-ionization current. It was numerically observed that up to 300
    TW/cm$^{2}$ the formation
    of LOHG is caused mainly by the bound-electron nonlinearity, while for
    higher intensities the tunnel-ionization current plays the dominant role.
    It was also shown that a high intensity of the pump does not necessary
    lead to efficient LOHG, rather, due to the reduced coherence length by
    the plasma contribution to the refraction index an optimum around
    100 TW/cm$^{2}$ with efficiencies  in the range of $3\times10^{-3}$ and $10^{-4}$
    for third and fifth harmonic generation, respectively,
    is predicted. Further on, we studied THG by intense pump pulses at 400 nm and predicted
    frequency transformation to the spectral range of 133 nm with maximum efficiency
    of about $1.5\times10^{-3}$.
    \section*{Acknowledgments} We acknowledge financial support by the German Research Foundation (DFG), project No. He $2083/17-1$.
    \end{document}